\documentclass[review]{elsarticle}

\usepackage{lineno,hyperref}
\usepackage{xspace}
\modulolinenumbers[5]
\usepackage{bm}
\usepackage{amssymb}
\usepackage{hyperref}
\usepackage{amsmath}
\usepackage{graphicx}

\journal{Journal of \LaTeX\ Templates}

\begin{document}

\begin{frontmatter}

\title{Universality of the underlying event in pp collisions}

\author{Antonio Ortiz}
\address{Instituto de Ciencias Nucleares, Universidad Nacional Aut\'onoma de M\'exico, \\ Apartado Postal 70-543,
M\'exico Distrito Federal 04510, M\'exico}

\author{Lizardo Valencia Palomo}
\address{Facultad de Ciencias en F\'isica y Matem\'aticas, Universidad Aut\'onoma de Chiapas, \\ Rancho San Francisco, Ciudad Universitaria, Tuxtla Guti\'errez, Chiapas}

\cortext[mycorrespondingauthor]{Corresponding author}
\ead{lizardo.valencia.palomo@cern.ch}

\begin{abstract}

In this paper we study ATLAS results on underlying event in pp collisions at $\sqrt{s}=0.9$, 7 and 13\,TeV. We show that the center-of-mass energy dependences of the charged-particle production sensitive to the underlying event (``transverse'' region) and to the hardest partonic interaction (``towards'' and ``away'' regions) in pp collisions can be both understood in terms of the change of the inclusive average multiplicity. Within uncertainties, the corresponding particle production  as a function of the leading charged particle shows no significant $\sqrt{s}$-dependence for the three regions once they are scaled according to the relative change in multiplicity.  The scaling properties reported here are well reproduced by PYTHIA 8.212 tune Monash 2013 and suggest an universality of the underlying event in hadronic interactions at high $\sqrt{s}$. Based on the simulations, we observed that the same scaling properties are also present in the average number of multi-partonic interactions as a  function of the leading charged particle. Moreover,  the multiplicity distributions associated to the underlying event exhibit a KNO scaling. 

\end{abstract}


\end{frontmatter}


\section{Introduction}

At the Large Hadron Collider (LHC) energies, the interacting objects are quarks and gluons (partons) which only carry a small fraction of the proton momentum ($x$). For instance, considering minimum bias proton-proton (pp) collisions at $\sqrt{s} = 13$ TeV, most of the partonic interactions at mid-rapidity are among gluons with $x\approx 2p_{\rm T}/\sqrt{s}\approx 10^{-3}$~\cite{Rojo:2014kta}. An accurate description of these processes is essential for simulating single pp interactions as well as the pile up effects that become important in high intensity runs at hadron colliders. Particle interactions at these energy scales are typically described by QCD-inspired models implemented in Monte Carlo (MC) event generators~\cite{Proceedings:2016tff}.

A typical pp collision can be visualised as a main hard partonic scattering surrounded by the so-called Underlying Event (UE). The UE can receive contributions not only from multiple partonic interactions and color reconnection (CR) between hard partons and beam remnants~\cite{Sjostrand:1987su}, but also from processes typically associated with hard physics like initial- and final-state radiation~\cite{Aad:2014hia}.  Clearly, it is difficult to isolate the UE component of the interaction, because even in the most naive model of particle production the interaction among colored objects before hadronization has to be considered. Such  observables  have already been  studied in pp collisions from $\sqrt{s} =$ 0.9 to 13\,TeV by the LHC experiments~\cite{Aad:2010fh,Aad:2011qe,Aad:2012jba,Aad:2014hia,Aad:2014jgf,Aaboud:2017fwp,ALICE:2011ac,Khachatryan:2010pv,Khachatryan:2015jza,Chatrchyan:2012tb}. Also, some measurements in p$\bar{\rm p}$ collisions in dijet and Drell-Yan events at CDF with center-of-mass energies of $\sqrt{s}=1.8$ and 1.96\,TeV~\cite{Affolder:2001xt,Acosta:2004wqa,Aaltonen:2010rm} are available.  

The study of the UE has become attractive also for the heavy-ion community, because now the the multiplicity dependent UE effects has to be separated from those attributed to new physics~\cite{Ortiz:2016kpz,Martin:2016igp}. At the LHC many interesting results have been obtained from pp data, in particular at high multiplicity pp collisions where the possible formation of a deconfined medium (sQGP~\cite{Mangano:2017plv}) is under discussion. Interestingly, most of the observables have been found to scale with the event multiplicity, rather than with the center-of-mass energy. Among them, the identified particle production reported by the CMS~\cite{Chatrchyan:2012qb} and the ALICE~\cite{ALICE:2017jyt} collaborations have unveiled a strong multiplicity dependence that makes the multiplicity dependent transverse momentum spectral shapes to be compatible with the expected effects driven by radial flow. In addition, the event shapes studies reported for the first time using minimum bias events exhibit a remarkable multiplicity dependence that is not described by the QCD-inspired MC generators~\cite{Abelev:2012sk,Ortiz:2017jho}. 

It also has to be reminded that high multiplicity events can be associated with small impact parameters, which can be selected by requiring a hard-$p_{\rm T}$ scale (``jet pedestal effect'')~\cite{Martin:2016igp}. Since for pp collisions the $\sqrt{s}$ increase is accompanied by an increase of the multiplicity~\cite{Adam:2015pza}, then the recent measurement on UE~\cite{Aaboud:2017fwp} should have an implicit multiplicity dependence that can be disentangled using the UE data at different $\sqrt{s}$. The understanding of the multiplicity dependent effects is important for the correct interpretation of the results. The main goal of the present paper is therefore to study whether we can factorize the multiplicity and $\sqrt{s}$-dependences of the UE-related observables in order to gain some insight into the physics mechanism behind the pp collisions. 

The article is organised as follows: section 2 outlines definitions of the observables measured in the underlying event analysis and introduces the data set used in the present study.  The scaling procedure is described  in  section  3 as well as results and comparisons with PYTHIA 8.212~\cite{Sjostrand:2014zea}.  Finally section 4 contains a summary and outlook.

\section{The underlying event observables}

\paragraph{Number density.} The particle density, expressed as the mean number of charged particles
\begin{equation}
\langle \frac{ N_{ch} }{\Delta \eta \Delta \phi} \rangle
\end{equation}
is  computed  as  the  average  number  of  primary  charged  particles  per unit  pseudorapidity ($\Delta \eta$) and  per  unit  azimuthal  separation ($\Delta \phi$).  The latter is the angular difference between  a  charged particle
and the leading charged particle, i.e. the largest transverse momentum of the event ($p_{\rm T}^{\rm leading}$ = hard process scale). The azimuthal angle $\phi$ lies  in the plane perpendicular to the beam axis. 

\begin{figure}[t!]
\begin{center}
\includegraphics[keepaspectratio, width=1.0\columnwidth]{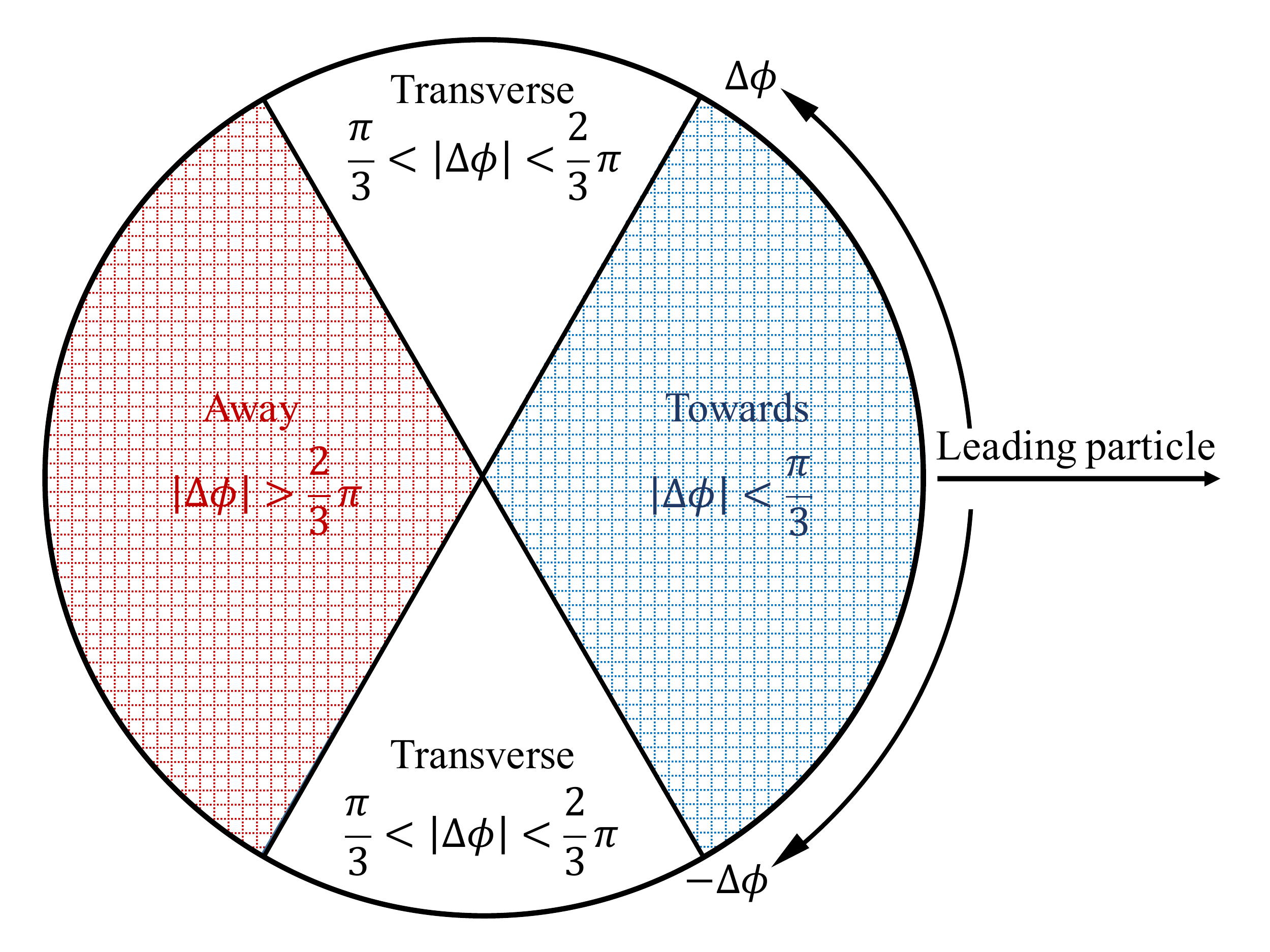}
\caption{\label{fig:0} (Color online). Definition of regions in the azimuthal angle with respect to the leading primary charged particle. The sensitivities in the  towards, away and transverse regions are described in the text.} 
\end{center}
\end{figure}

\paragraph{Summed transverse momentum.}   The mean scalar $p_{\rm T}$ sum per unit pseudorapidity and per unit azimuthal
separation
\begin{equation}
\langle \frac{ \sum_{i} p_{\rm T,i} }{\Delta \eta \Delta \phi} \rangle
\end{equation}
 is obtained using  the transverse momenta of primary charged particles. Primary charged particles are defined as those with a mean lifetime $\tau>300$\,ps, which are either directly produced in the pp interaction or from subsequent decays of particles with a lifetime $\tau<30$\,ps.

This measurement uses the established form of UE observables~\cite{Field:2002vt}, with the azimuthal plane of the event segmented into several distinct regions having different sensitivities to the UE. As illustrated in Fig.~\ref{fig:0}, the azimuthal angular difference with respect to the leading charged particle, is used to define three regions.
\begin{itemize}
\item{``Towards'' region}: $|\Delta \phi|<\frac{\pi}{3}$.
\item{``Transverse'' region}: $\frac{\pi}{3}<|\Delta \phi|<\frac{2}{3}\pi$.
\item{``Away'' region}: $|\Delta \phi|>\frac{2}{3}\pi$.
\end{itemize}

\begin{figure*}[htbp]
\begin{center}
   \includegraphics[width=0.9\textwidth]{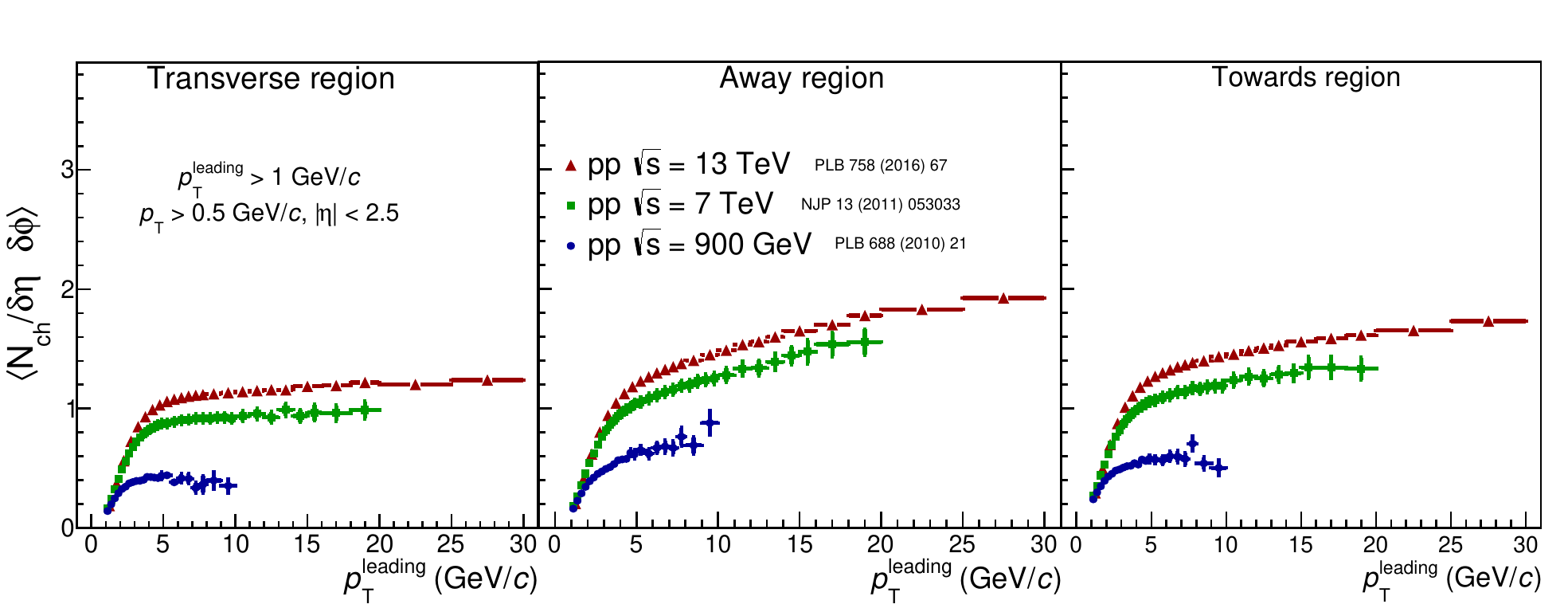}
   \includegraphics[width=0.9\textwidth]{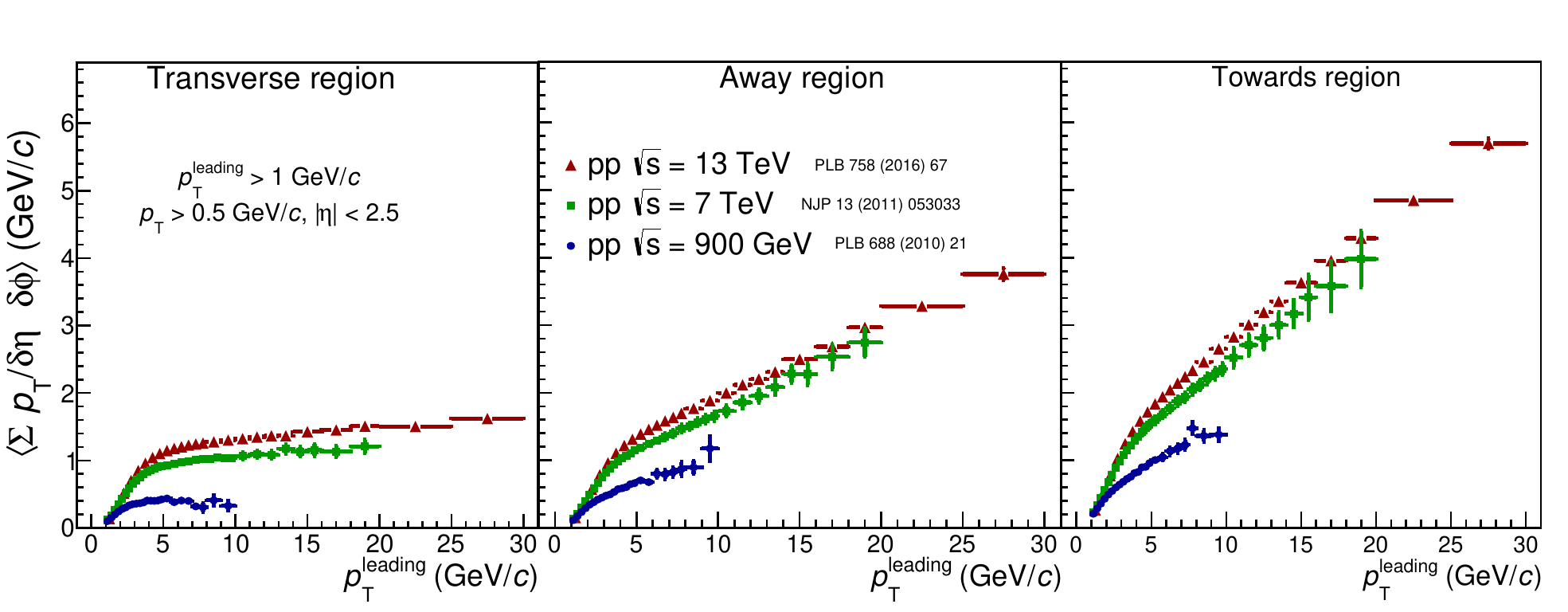}   
   \caption{\label{fig:1} (Color online).  Number density of primary charged particle multiplicity (top) and the mean scalar $p_{\rm T}$ sum (bottom) as a function of the transverse momentum of the leading charged particle. Measurements were performed by the ATLAS Collaboration using primary charged particles in pp collisions at $\sqrt{s}=0.9$, 7 and 13\,TeV~\cite{Aad:2010fh,Aaboud:2017fwp}. Results for the transverse (left), away (middle) and towards (right) regions are displayed.}
\end{center}
\end{figure*}

As the scale of the hard scattering increases, the leading charged particle acts as a convenient indicator of the main flow of the hard-process energy.  The towards and away regions are dominated by particle production from the hard process. In contrast, the transverse region is more sensitive to the UE and observables measured inside it are the primary focus of UE measurements. The variables calculated in this analysis are constructed using primary charged particles with $p_{\rm T}>0.5$\,GeV/$c$ and $|\eta|<2.5$ with the leading charged particle requirement of $p_{\rm T}^{\rm leading}>1$\,GeV/$c$.

Figure~\ref{fig:1} shows the energy dependence of the number densities of charged-particle multiplicity (top panel) and the mean scalar $p_{\rm T}$ sum (bottom panel) as a function of $p_{\rm T}^{\rm leading}$ in the transverse, towards, and away regions. The data points were obtained from the ATLAS Collaboration~\cite{Aad:2010fh,Aaboud:2017fwp}. It is worth noting that the slopes of the number densities as a function of $p_{\rm T}^{\rm leading}$ seem to be $\sqrt{s}$-independent for $p_{\rm T}^{\rm leading}>4$\,GeV/$c$. A similar behavior is also observed for the summed $p_{\rm T}$. The potential scaling properties of these observables are explored in the next section.

\section{Results and discussion}

The underlying event analysis considers primary charged particles with transverse momentum above a given threshold. In addition, depending on the features of the experiments different intervals of pseudorapidity are used. In the present paper we consider the selection at the particle level from ATLAS ($p_{\rm T}>0.5$\,GeV/$c$ and $|\eta|<2.5$), however, the conclusions should hold for narrower pseudorapidity intervals~\cite{ALICE:2011ac}. In order to obtain the average multiplicity value at $|\eta|<2.5$, the transverse momentum distributions of inclusive charged primary particles in pp collisions at $\sqrt{s}=0.9$, 7 and 13\,TeV~\cite{Aad:2010rd,Aad:2010ac,Aad:2016mok} were integrated for $p_{\rm T}>0.5$\,GeV/$c$. The total uncertainties (quadratic sum of statistical and systematic uncertainties) associated to the spectrum were propagated to the integral and assigned as systematic uncertainty to the average multiplicity.  For pp collisions at $\sqrt{s}=0.9$, 7 and 13 TeV, we obtained $\langle {\rm d}N_{\rm ch}/{\rm d}\eta\rangle_{|\eta|<2.5, \,p_{\rm T}>0.5\,{\rm GeV/}c}=1.306 \pm 0.069$, $2.405 \pm  0.070$ and $2.862 \pm 0.037$, respectively. 

\begin{figure*}[htbp]
\begin{center}
   \includegraphics[width=0.9\textwidth]{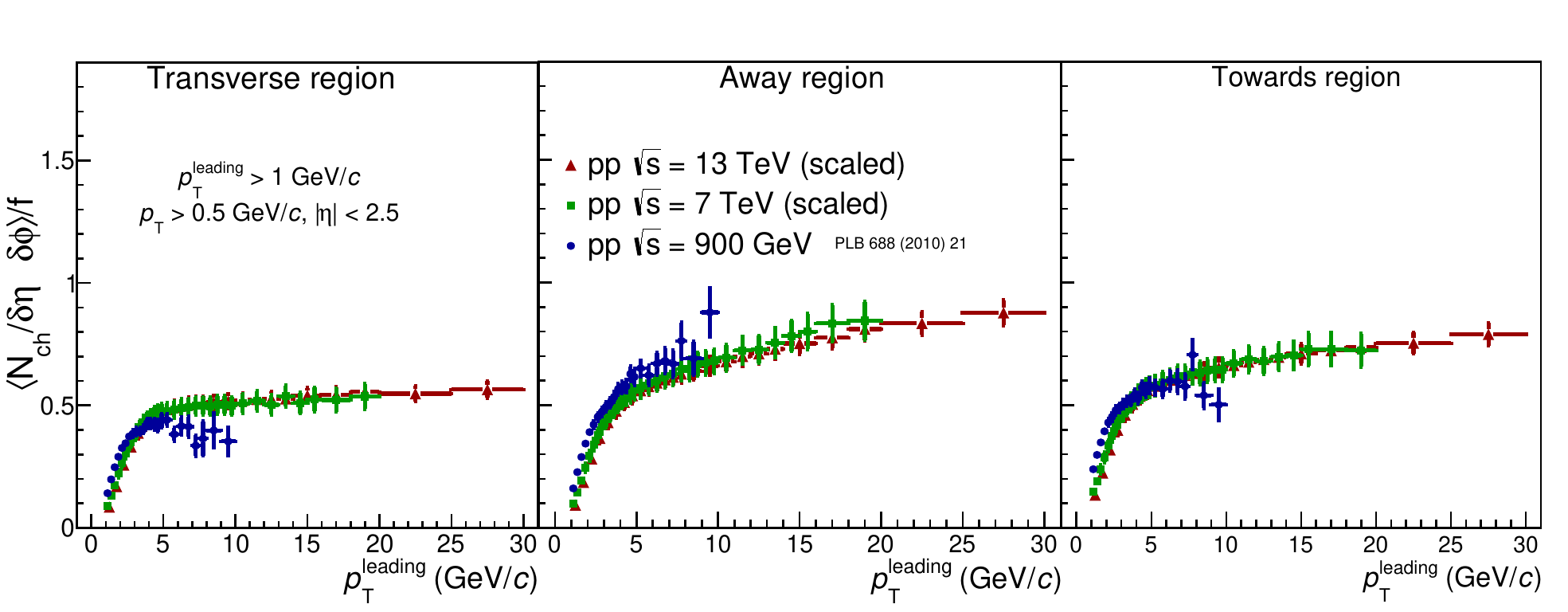}
   \includegraphics[width=0.9\textwidth]{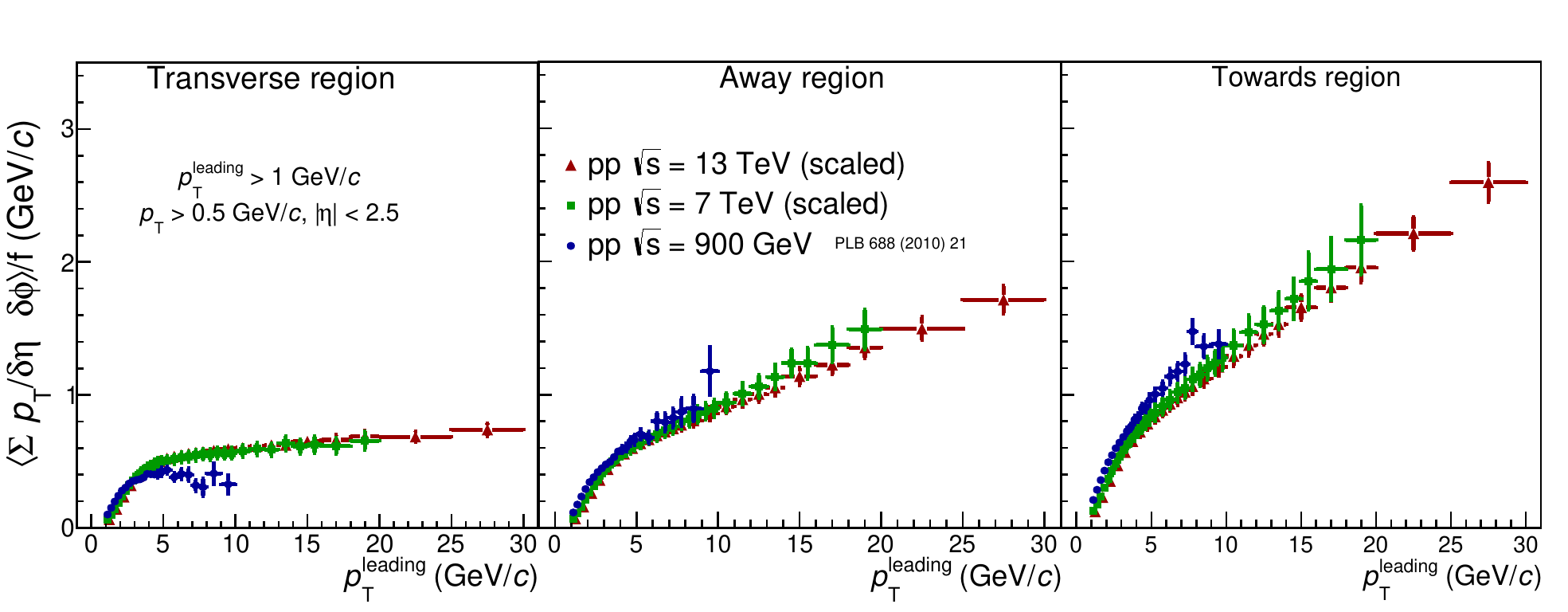}   
   \caption{(Color online).  The mean densities of charged-particle multiplicities and the mean scalar $p_{\rm T}$ sum as a function of the transverse momentum of the leading particle are shown in the top and bottom panel, respectively. A factor $f$ which takes into account the relative multiplicity increase from 0.9 to 13 TeV was applied to those quantities. Results for the transverse, away and towards regions are displayed in the left, middle and right panel, respectively. Error bars include the systematic uncertainty from the scaling factor and the total uncertainties reported by ATLAS.}
  \label{fig:2}
\end{center}
\end{figure*}

Using those values, the relative variation ($f$) of the average multiplicity with respect to pp collisions at $\sqrt{s}=0.9$\,TeV was calculated. The number density and the summed $p_{\rm T}$ were both scaled considering the factors $f=1$, 1.842 and 2.191 for pp collisions at $\sqrt{s}=0.9$, 7 and 13\,TeV, respectively. Figure~\ref{fig:2} shows the number densities (top panel) and the summed $p_{\rm T}$ (bottom panel) for the three regions considered after scaling by $1/f$ factor. In this representation these quantities show, within uncertainties, little center-of-mass energy dependence.  For the towards and the away regions, we observe that the multiplicity density increases monotonically with the $p_{\rm T}^{\rm leading}$ scale. While in the transverse region, after a monotonic increase at low $p_{\rm T}^{\rm leading}$ ($<5$\,GeV/$c$), the distribution tends to flatten out. This is the so-called ``jet pedestal effect'' which can be interpreted as the evidence for an impact parameter dependence in the hadronic collision~\cite{Sjostrand:1987su} that makes the bulk particle production independent of the hard scale.

It is worth noting that the same factor $f$, which takes into account the change in the multiplicity associated to soft physics, is able to unveil these scaling properties for regions having different sensitivities to the underlying event. The small $\sqrt{s}$-dependence observed in data can be understood as follows. If the scaling holds for the transverse region (as supported by data due to its large sensitivity to soft physics), then the UE components within the towards and away sides are expected to scale with the event multiplicity. Since particle production associated to jets is $\sqrt{s}$-independent (jet universality), then the small $\sqrt{s}$-dependence observed in the towards and away regions could be attributed to different contributions from quark and gluon jets. 

It is important to remaind that ALICE has reported the $\sqrt{s}$-dependence of the particle production in the so-called plateau interval~\cite{ALICE:2011ac}; $p_{\rm T}^{\rm leading}$-range from 4 to 10\,GeV/c at $\sqrt{s}=0.9$\,TeV and from 10 to 25\,GeV/$c$ at $\sqrt{s}=7$\,TeV both in the transverse region (UE). The relative increase of the activity in that region from 0.9 to 7 TeV was found to be larger than that for the ${\rm d}N_{\rm ch}/d\eta$. Such an observation is consistent with what we see in our study.  

\begin{figure*}[htbp]
\begin{center}
   \includegraphics[width=0.9\textwidth]{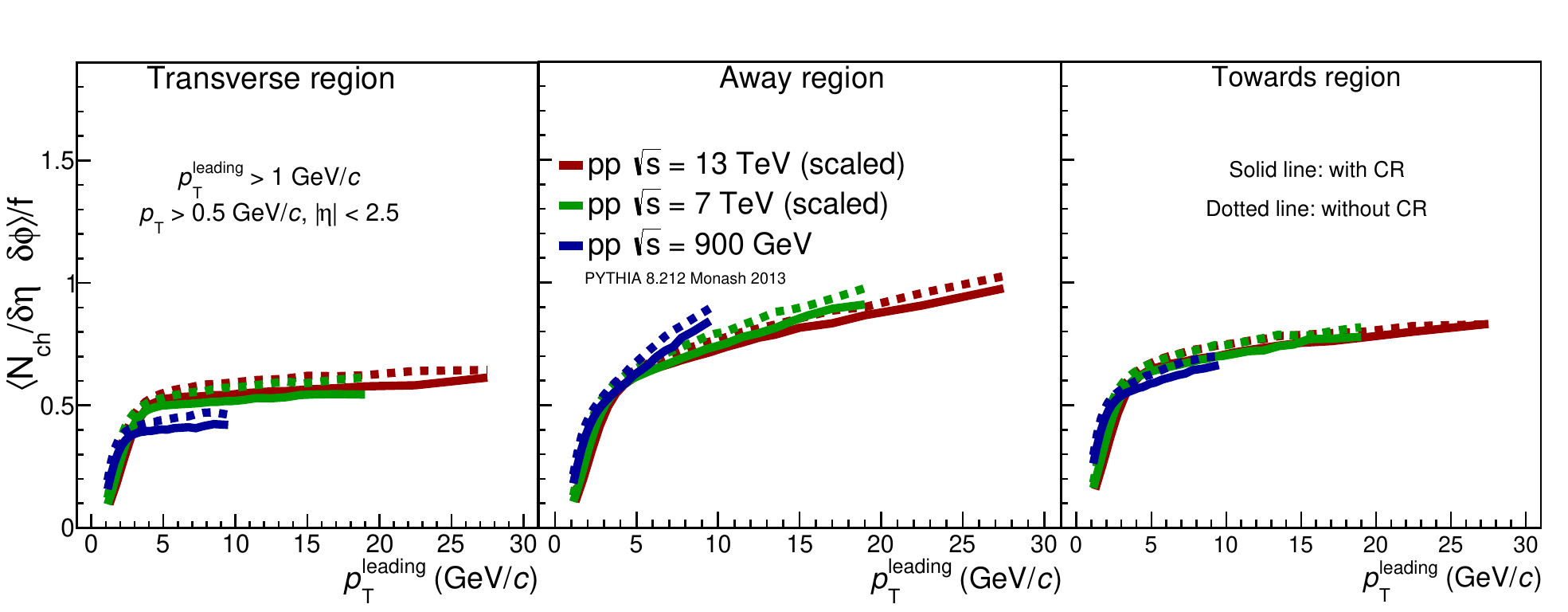}
   \includegraphics[width=0.9\textwidth]{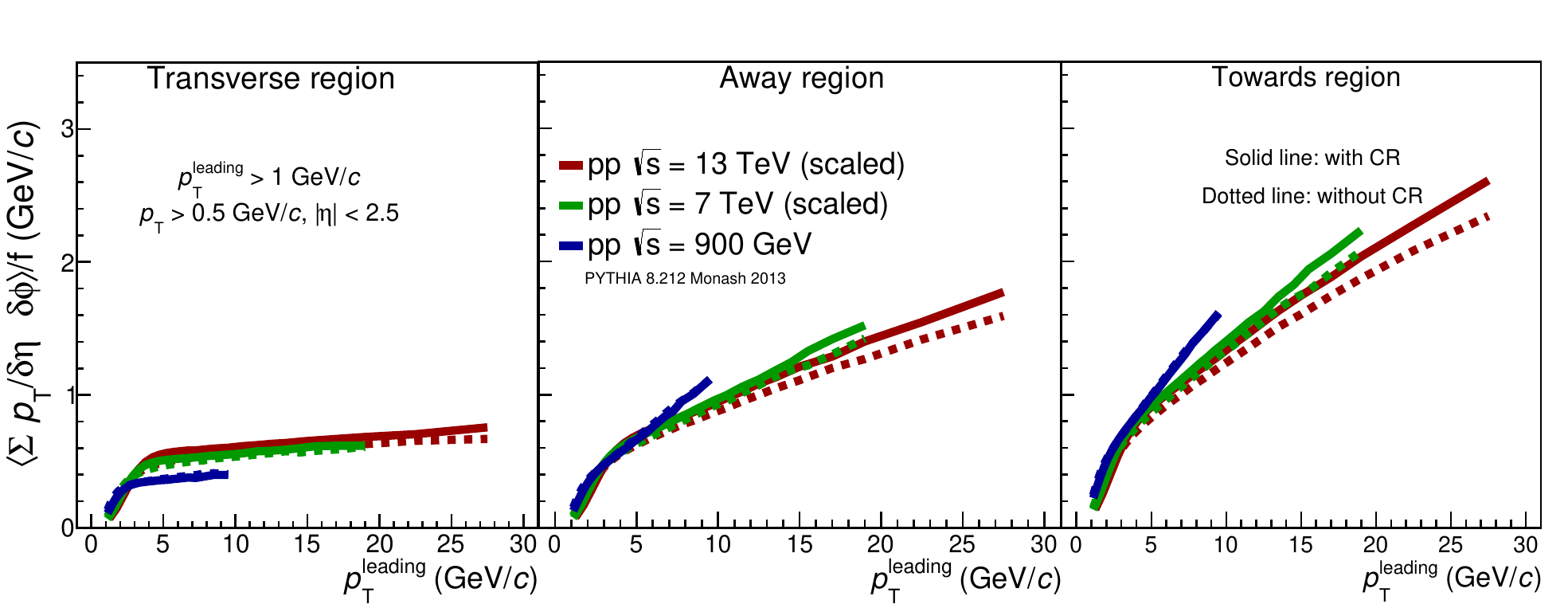}   
   \caption{(Color online).   Proton-proton collisions simulated with PYTHIA 8.212. The mean densities of charged-particle multiplicities and the mean scalar $p_{\rm T}$ sum as a function of the transverse momentum of the leading particle are shown in the top and bottom panel, respectively. A factor $f$ which takes into account the relative multiplicity increase from 0.9 to 13 TeV was applied. Results for the transverse, away and towards regions are displayed in the left, middle and right panel, respectively.}
  \label{fig:3}
\end{center}
\end{figure*}

To complement the study, Fig.~\ref{fig:3} shows the multiplicity-scaled number density and summed $p_{\rm T}$ for pp collisions at $\sqrt{s}=0.9$, 7 and 13\,TeV simulated with  PYTHIA 8.212~\cite{Sjostrand:2014zea}  tune Monash 2013~\cite{Skands:2014pea}. Surprisingly, the scaling is also observed in MC simulations. For UE studies, color reconnection can play an important role in the observed behavior because it models the interaction between partons of different $p_{\rm T}$ scales just before the hadronization.  It is also  important for reproducing the rise of the average transverse momentum with the event multiplicity~\cite{Abelev:2013bla}, as well as radial flow-like patterns in high multiplicity events~\cite{Ortiz:2013yxa}. 

In order to investigate potential color reconnection effects on the underlying event quantities, Fig.~\ref{fig:3} also shows a comparison of results when color reconnection is switched on or off. Clearly, the scaling is also observed in simulations without CR, however, the results exhibit interesting features. On one hand, at $\sqrt{s}=7$ and 13\,TeV CR increases the summed $p_{\rm T}$ with increasing $p_{\rm T}^{\rm leading}$ in the three regions; on the other hand, it reduces the number density (multiplicity) for the three energies. This behavior is a manifestation of the interplay between hard (large $p_{\rm T}^{\rm leading}$) and soft physics (summed $p_{\rm T}$ considering low momentum charged particles) which increases the particle's $p_{\rm T}$ via color strings connecting low momentum partons with others of higher $p_{\rm T}$ scales.

Another posible cause of the observed effect are the multi-partonic interactions (MPI) because in PYTHIA this is the mechanism which produces the UE. Therefore, we  investigated whether the factor $f$ has some sensitivity to the change of the average number of multi-partonic interactions ($\langle N_{\rm mpi} \rangle$) going from low to high $\sqrt{s}$.  

\begin{figure}[t!]
\begin{center}
\includegraphics[keepaspectratio, width=1.0\columnwidth]{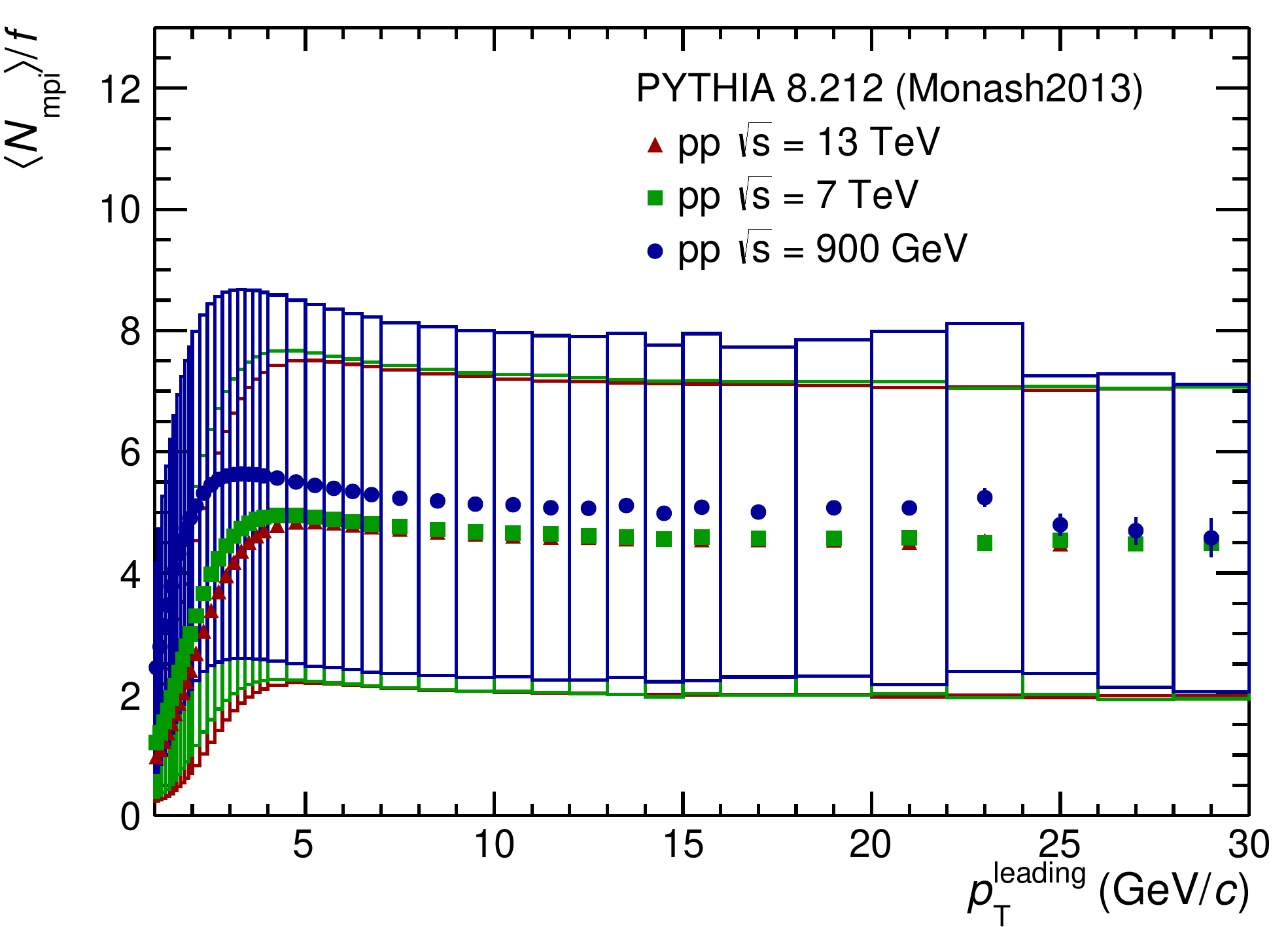}
\caption{\label{fig:5} (Color online). Average number of multi-partonic interactions ($\langle N_{\rm mpi} \rangle$), scaled by $f$, as a function of the transverse momentum of the leading particle. The factor $f$ was calculated using the inclusive charged particle production within $|\eta|<2.5$. The boxes around the points indicate the RMS of the $N_{\rm mpi}$ distributions for each momentum interval. Results for pp at $\sqrt{s}=0.9$, 7 and 13\,TeV are displayed.} 
\end{center}
\end{figure}

Figure~\ref{fig:5} shows the average number of multi-partonic interactions, scaled by $f$, as a function of the transverse momentum of the leading charged particle. First of all, the figure illustrates the impact parameter dependence of the UE, i.e. going from low to high $p_{\rm T}^{\rm leading}$ the average number of MPI increases up to $p_{\rm T}^{\rm leading}\sim4$\,GeV/$c$ and then it remains constant because the ``collision centrality'' gets fixed. Second, in the jet pedestal region, $p_{\rm T}^{\rm leading}>4$\,GeV/$c$, the average number of MPI exhibits the same scaling properties which were observed in the transverse region. In other words, within 10\% the relative change in the number of multi-partonic interactions (going from $\sqrt{s}=0.9$\,TeV to higher energies) agrees with the relative change in the average multiplicity calculated using charged particles within $|\eta|<2.5$ and $p_{\rm T}>0.5$\,GeV/$c$. It is important to say that the usage of minimum bias pp events for the calculation of the charged particles multiplicity played a crucial role because this quantity is very sensitive to MPI.  

The scaling properties reported in this paper suggest a sort of KNO scaling~\cite{Koba:1972ng} of the underlying event. The KNO scaling is the hypothesis that at high energies $\sqrt{s}$ the probability distributions $P(n)$ of producing $n$ particles in a certain collisions process should exhibit the scaling relation:
\begin{equation}
P(n)=\frac{1}{\langle n \rangle} \Psi (\frac{n}{\langle n \rangle})
\end{equation}
with $\langle n \rangle$ being the average multiplicity. This means that the rescaled data points $P(n)$ measured at different energies collapse onto the unique curve $\Psi$~\cite{Hegyi:2000sp}. An analysis of the multiplicity distributions of pp and p${\bar{\rm p}}$ data from $\sqrt{s}=23.6$\,GeV to $\sqrt{s}=1.8$\,TeV shows that the KNO scaling does not hold except for very central and small regions of phase space~\cite{GrosseOetringhaus:2009kz}. 

\begin{figure}[t!]
\begin{center}
\includegraphics[keepaspectratio, width=1.0\columnwidth]{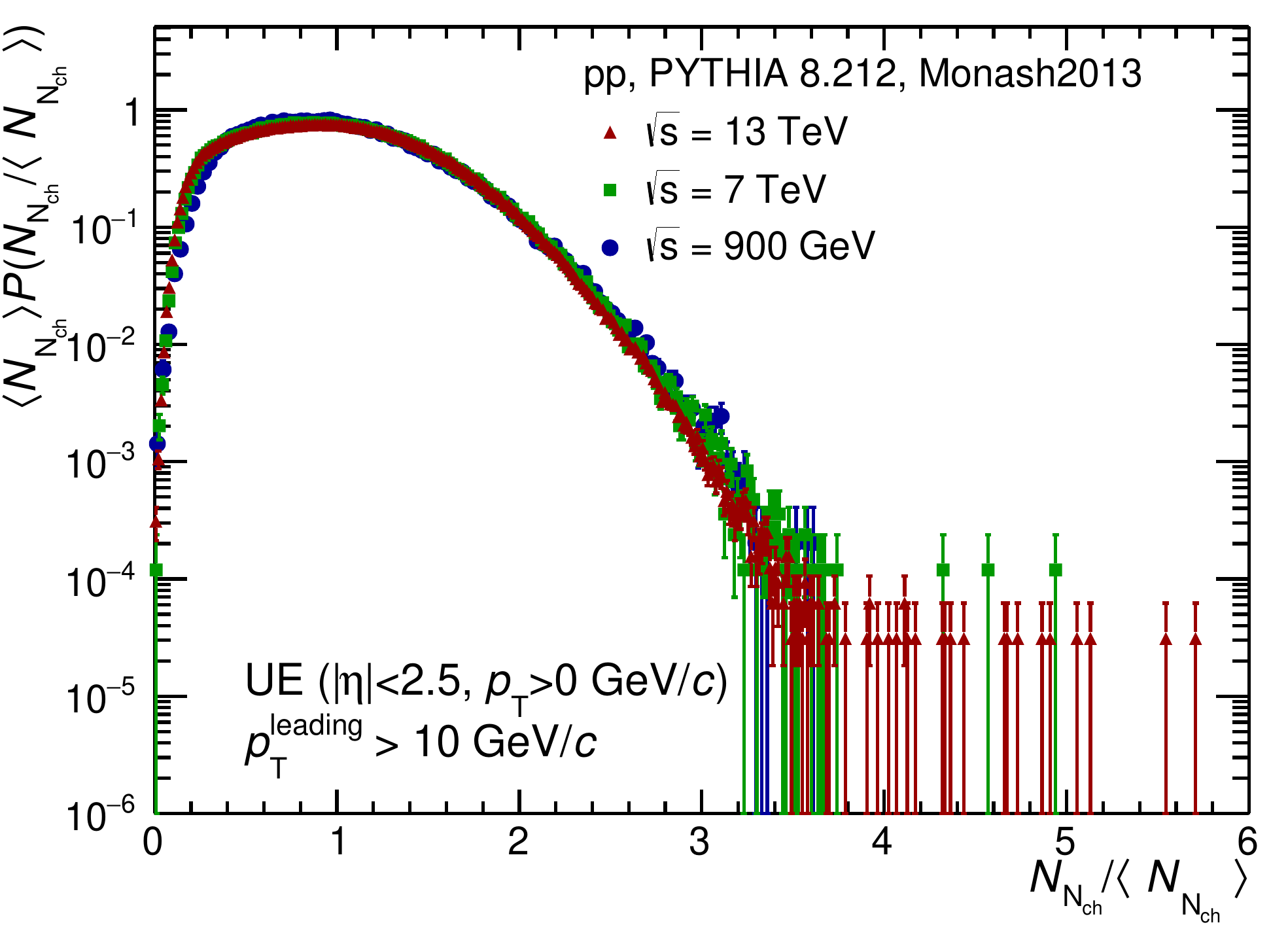}
\caption{\label{fig:4} (Color online). Charged particles multiplicity distributions in KNO variables of inelastic pp collisions simulated with PYTHIA 8.212. Results for different collisions energies are shown for the UE region.} 
\end{center}
\end{figure}

To check whether the UE exhibits a KNO scaling, the charged particles multiplicity distributions ($|\eta|<2.5$, $p_{\rm T}>0$) of the UE region (transverse region) are studied in KNO variables. Figure~\ref{fig:4} shows the results for pp collisions at $\sqrt{s}=0.9$, 7 and 13\,TeV simulated with PYTHIA 8.212 and requiring the leading particle to satisfy $p_{\rm T}^{\rm leading}\geq10$\,GeV/$c$. In this representation all the distributions seem to collapse onto the same universal curve. A similar effect was already reported for the so-called rare events $C$~\cite{DiasdeDeus:1997ui}; assuming that a single hadron-hadron collision results from the superposition of $\nu$ elementary partonic collisions emitting independently, it can be shown that the probability distributions associated to a rare event $C$ obeys the universal relation:
\begin{equation}
P_{C}(\nu)=\frac{\nu}{\langle \nu \rangle} P(\nu)
\end{equation}
Which is exactly what we get assuming that $N_{\rm ch}$ is proportional to $\nu$, being $\nu$ identified with $N_{\rm mpi}$.
The KNO scaling properties could be further checked in data  by studying multiplicity distributions associated to either jets or underling event. For this reason, the event shape analysis has been proposed as a tool to isolate jet and UE enriched samples in pp collisions~\cite{Ortiz:2017jho}. 

In summary, for the first time we have shown an universality of the underlying event for high energy pp collisions. The scaling behaviour reported in this paper can provide an estimation of the underlying activity at higher colliding energies. For example at $\sqrt{s} = 14$ TeV, the designed top LHC energy, or even for long term projects like the Future Circular Collider aimed to collide protons at $\sqrt{s} = 100$ TeV~\cite{Mangano:2016jyj}.

\section{Conclusion}

In this paper we have used ATLAS measurements on underlying event in pp collisions at $\sqrt{s}=0.9$, 7 and 13\,TeV. We have found that the energy dependence of such quantities as a function of the hard transverse momentum scale can be explained as a consequence of the change in the average multiplicity density of primary charged particles. The study was conducted considering charged particles with $p_{\rm T}>0.5$\,GeV/$c$ and $|\eta|<2.5$. Such a scaling property was observed for three regions sensitive to different processes: towards and away (sensitive to the fragmentation products of the hardest partonic interaction), and transverse (the most sensitive to the underlying event activity) regions. Since the scaling is reproduced by PYTHIA 8.212, we performed an analysis and found that the same scaling is unveiled if one considers the average number of multi-partonic interactions instead of charged particles multiplicity or summed $p_{\rm T}$. Indeed, for $p_{\rm T}^{\rm leading}\geq10$\,GeV/$c$ the relative change in the number of multi-partonic interactions (going from $\sqrt{s}=0.9$\,TeV to higher energies) agrees (within 10\%) with the relative change in the average multiplicity calculated using charged particles within $|\eta|<2.5$ and $p_{\rm T}>0.5$\,GeV/$c$. In addition,  we found that multiplicity distributions in the underlying-event region exhibit a KNO scaling which is expected for a single pp collision involving a large number multi-partonic interactions emitting independently. The results support the existence of MPI and allows us to learn about the geometry in pp collisions. Therefore, the results presented here may  serve  as  a  useful  input  to understand the new data provided by the LHC experiments, in particular the new phenomena discovered in high multiplicity pp collisions.

\section{Acknowledgments}
We acknowledge Peter Christiansen, Andreas Morsch and Guy Pai{\'c} for the useful comments and discussions. We also acknowledge the technical support of Luciano Diaz and Eduardo Murrieta for the maintenance and operation of the computing farm at ICN-UNAM and also to the LARCAD-UNACH. Support for this work has been received from CONACyT under the grant number 280362.

\section*{References}

\bibliography{mybibfile}

\end{document}